\documentstyle[prd,aps,epsfig]{revtex}
\newcommand{\eq}{\begin{equation}}
\newcommand{\beqn}{\begin{eqnarray}}
\newcommand{\en}{\end{equation}}
\newcommand{\eeqn}{\end{eqnarray}}
\def\part{\partial}

\def\uel{u_{\Lambda=1 \, \ell  m}}
\def\intp{\int_{-\pi/2}^{\pi /2}}
\def\ci{{\rm ci}}
\def\Si{{\rm Si}}
\begin{document}
\draft
\preprint{UIUC-98/2; HUTP-98/A010; gr-qc/9803073}
\title{ Where do all the Supercurvature Modes Go? }
\author{J. D. Cohn$^{(1)}$ and D. I. Kaiser$^{(2)}$}
\vskip 1cm
\address{$(1)$ Departments of Physics and Astronomy, 
University of Illinois \\
Urbana-Champaign, IL 61801 \\
{\rm jdc@uiuc.edu} \\
\bigskip
$(2)$  Department of Physics, 
Harvard University \\
Cambridge, MA 02138 \\
{\rm dkaiser@fas.harvard.edu} \\}
\date{March 1998}
\maketitle

\begin{abstract}
In the hyperbolic slicing of de Sitter space appropriate for open
universe models, a curvature scale is present and
supercurvature fluctuations are possible.
In some cases, the
expansion of a scalar field in the Bunch-Davies vacuum includes
supercurvature modes, as shown
by Sasaki, Tanaka and Yamamoto. 
We express the normalizable vacuum 
supercurvature modes for a massless scalar field in terms of the basis modes
for the spatially-flat slicing of de Sitter space.  
\end{abstract}
\pacs{Preprint UIUC-98/2; \hskip 0.5 cm HUTP-98/A010; \hskip 0.5 cm
gr-qc/9803073}
\vskip 2pc
%%%%%%%%%%%%%%%%%%%%%%%%%%%%%%%
\baselineskip 18pt

\section{Introduction}

Scalar fields in de Sitter spacetime have long provided a testing ground
for issues of quantum field theory in curved spacetime
\cite{birdavies,sfulling}.  Further motivation for their study stems from
the central role they play in inflationary cosmology \cite{KolbTurner}.
Several different coordinate systems can be used to cover de Sitter
space, and subtleties in the quantization of fields can arise in some of
the less
familiar coordinate systems.  These subtleties have been highlighted by
recent models of open inflation \cite{bgt,openi}, in which two periods of
inflation are separated by nucleation of a bubble.  The bubble interior
includes an open universe ($\Omega_0 <1$), where we could be living
today, described by hyperbolic, spatially-curved coordinates
\cite{colegott}.

A key difference between these hyperbolic coordinates and the more
familiar spatially-flat slicing of de Sitter space is the presence of a
curvature scale.  This in turn leads to the possibility of supercurvature
\cite{lythwos} fluctuations, fluctuations with wavelength longer than the
curvature scale.  Unlike the continuum of modes familiar from the
spatially-flat slicing of de Sitter space, a normalizable supercurvature
mode may exist for an isolated, discrete eigenvalue of the spatial
Laplacian, or not at all.  Although such modes have no analogue in the
spatially-flat slicing of de Sitter space, it has been shown
\cite{sasaki95} (see also \cite{Mosch}) 
that the supercurvature modes must be included in
the vacuum spectra of low-mass scalar fields in order to produce a
complete set of states, and hence the proper Wightman function. 

For massless, minimally-coupled scalar fields in de Sitter space, there
is
in addition a well-known infrared divergence in the
(coordinate-independent) Wightman function.  The infrared divergence is
related to a dynamical zero mode in the spectrum of a massless field
\cite{desinfra,kg93}.  Kirsten and Garriga \cite{kg93} covariantly
quantized this zero mode in a spatially-closed slicing of de Sitter
space.
Extending their result from these closed coordinates to the coordinate
system appropriate to open inflation has not yet been done.
Here we identify the zero mode as one of the supercurvature modes
when quantizing the minimally-coupled massless field in the open
hyperbolic coordinates.

It has already been noted that
supercurvature modes can make significant contributions to the
fluctuations
in the cosmic microwave background (CMB) radiation, and 
several of their effects have
been calculated for models of open inflation
\cite{sccmb,garriga,sasaki96,gar-97}.    (The massless
field zero-mode subtleties, except for a variant studied
in \cite{gar-97}, do not arise 
for these specific CMB calculations, which are sensitive to higher
multipoles.)
In addition to contributing to observable density fluctuations, 
such long wavelength, supercurvature modes might play a role \cite{dkpre}
at the end of inflation, when recent advances \cite{newreh}
in the theory of reheating are taken into consideration.

\indent  In summary, increased understanding 
of these supercurvature modes is motivated both by general
questions of quantizing fields in curved backgrounds, and by recent
inflationary model-building.   We will focus here on the 
example of supercurvature modes for a massless,
minimally-coupled scalar field, expressing them
as a sum over the basis modes for a spatially-flat slicing of de Sitter
space.  This overlap gives a measure of
\lq\lq where all the supercurvature
modes go" in the familiar flat-space spectrum of such fields.  
The massless case is chosen for tractability, and questions about the
zero mode are postponed for future work \cite{jdcdk}.

Throughout this paper, we
consider only an unperturbed de Sitter metric; a more complete treatment
would include study of the backreaction of such fields on the background
metric.  Because the supercurvature modes stretch beyond the horizon, any
such study of the coupled metric fluctuations would need to pay special
attention to the gauge subtleties which always accompany superhorizon
fluctuations \cite{rhb},
and such issues are not pursued here. 

In section II, the two covers of de Sitter space (including the flat
and hyperbolic slicings) are given, and the field quantization pertinent
to open inflation in both systems is
reviewed.  As these supercurvature modes have
no analogue in the usual flat slicing of de Sitter space, this section
gathers some previous work on supercurvature
modes and provides notation and context for the rest of the paper.
Section III specializes to the massless case.  The
explicit calculation of the overlap for some of these supercurvature modes and
the more familiar flat space modes is given, and indicates a general
form for the overlap between all the
massless supercurvature and spatially flat modes.
We verify this general form by integrating over the
spatially flat modes, weighted by the overlap, to
obtain the original supercurvature modes.
Concluding remarks follow in Section IV.  Three Appendices include 
the supercurvature mode normalization at fixed time in the flat
coordinates, and details
of the overlap calculation along different hypersurfaces 
within de Sitter space.

\section{De Sitter Spacetime and Scalar Field Quantization}

We begin by considering de Sitter spacetime as embedded within
a 5-dimensional Minkowski spacetime, with the five coordinates subject to
the constraint \cite{birdavies}:
%%%%%%%%%%%%%%%
\eq
- \left( z^0 \right)^2 + \sum_{i = 1}^4 \left(z^i \right)^2 = 1.
\label{embed}
\en
The radius of the embedded spacetime, $H^{-1}$, is scaled to
unity.

A spatially flat slicing (see, {\it e.g.} \cite{birdavies}) 
which partially covers the resulting 4-dimensional de Sitter
spacetime is 
%%%%%%%%%%%%%%%
\beqn
\nonumber
z^0 &=& \sinh t_f + \frac{1}{2} e^{t_f} r_f^2 = - \frac{1}{2\eta} 
\left( 1 + r_f^2 \right) + \frac{1}{2} \eta , \\
\nonumber z^4 &=& \cosh t_f - \frac{1}{2} e^{t_f} r_f^2 = - \frac{1}{2\eta} 
\left( 1 - r_f^2 \right) - \frac{1}{2} \eta , \\
\vec{z} &=& e^{t_f} r_f \vec{\Omega}= - \frac{r_f}{\eta}\vec{\Omega} ,
\label{c1xflat2}
\eeqn
where $\vec{\Omega} \equiv (\sin \theta \cos \varphi,\> \sin
\theta \sin \varphi, \> \cos \theta)$,  
and conformal time
$d \eta = a^{-1} dt_f$, or $\eta = -a^{-1}(t_f)
= -e^{-t_f}$.  Here
$r_f \geq 0$,
$-\infty < \eta \leq 0$.  (See Figure 1.)  
%%%%%%%%%%%%%%%%%
\begin{figure}
\centerline{\epsfig{file=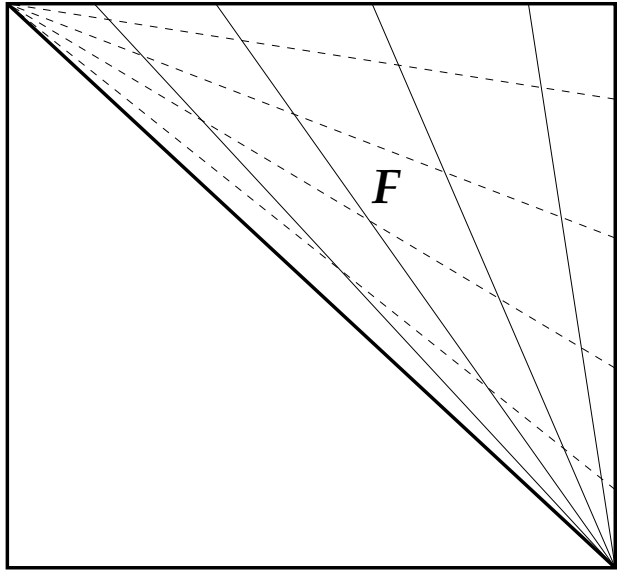,height=3.0in}}
\vspace{10pt}
{\caption{\small De Sitter spacetime as covered by the coordinates in
equation (\ref{c1xflat2}).  Solid lines are lines of constant $r$, and
dashed lines are lines of constant $\eta$.}}
\end{figure}
The metric on this
portion of the spacetime is
%%%%%%%%%%%%%%
\eq
ds_f^2 = \eta^{-2} \left[ - d\eta^2 + dr_f^2 + 
r_f^2 d\Omega^2 \right],
\label{c1dsF}
\en
where $d\Omega^2 \equiv d\theta^2 + \sin^2 \theta d\varphi^2$.  
The coordinates $x_f$ and
metric $ds_f^2$ are useful in ordinary models of
inflation, in which the spatial curvature quickly becomes completely
negligible.  
These coordinates $x_f$ cover only that half of the total spacetime 
with $z^0 +z^4 \geq 0$.
Replacing $z^4 \to - z^4$ in the above produces a
second flat coordinate system.  

In contrast, for models of open inflation,
open hyperbolic coordinates are appropriate inside the open
universe.  This coordinate patch is part of the full
de Sitter spacetime (discussed in detail by \cite{sasaki95,ballen})
as shown in Figure 2.
%%%%%%%%%%%%%%%%%%%%%%%%%%%%
\begin{figure}
\centerline{\epsfig{file=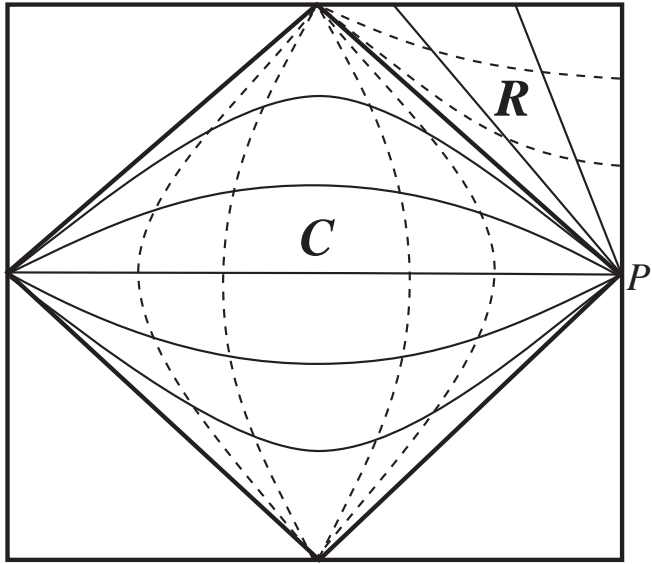,height=3.0in}}
\vspace{10pt}
\caption{\small De Sitter spacetime as covered by the coordinates in
equation (\ref{c1xopen}).  Solid lines are lines of constant $r$, and
dashed lines are lines of constant $t$.  If a (thin-walled) bubble
nucleated with center at point $P$, 
then region $R$, the forward light cone of point
$P$, would contain in the interior of the nucleated bubble. In models of
open inflation, region $R$ would contain our observable universe
at some time $t$ inside the light cone corresponding to $t=0$.}
\end{figure} 
The two most important of these regions for
our purposes here are region $C$, a large, compact subspace, and region
$R$, 
an open hyperboloid bordering region $C$.  These are related to the 
embedding coordinates by
%%%%%%%%%%%%%%%%
\beqn
\nonumber \; z^0 &=& \cos t_c \sinh r_c = \sinh t_r \cosh r_r , \\
\nonumber \; z^4 &=& \sin t_c = \cosh t_r , \\
\vec{z} \;  &=&  \cos t_c \cosh r_c \vec{\Omega} = \sinh t_r \sinh r_r
\vec{\Omega} .
\label{c1xopen}
\eeqn
These coordinates lie in the ranges
%%%%%%%%%%%%%%%%
\eq
- \frac{\pi}{2} \leq t_c \leq \frac{\pi}{2} \>\>,\>\> - \infty < r_c <
\infty \>\>{\rm and} \; \>\> 0 \leq t_r  \>\>,\>\> 0 \leq r_r ,
\en
and are related by the analytic continuation
$t_r = i t_c - {i\pi \over 2}$ and $r_r = r_c + {i\pi \over 2}$.  The
corresponding metrics are
%%%%%%%%%%%%%%%%%%%
\beqn
\nonumber ds_c^2 &=& dt_c^2 + \cos^2 t_c \left[ - dr_c^2 + \cosh^2 r_c
d\Omega^2 \right] , \\
ds_r^2 &=& - dt_r^2 + \sinh^2 t_r \left[ dr_r^2 + \sinh^2 r_r d\Omega^2
\right] .
\label{c1openmetrics}
\eeqn
Note that $r_c$ plays the role of \lq time' within region $C$.

The forward light cone of the point $P$ in Figure 2 is
the center of the nucleated bubble in models of open inflation,
so that region $R$ contains the spatially-open universe we may be
in today.   Region $C$ is of interest because surfaces of fixed time, 
$r_c = {\rm constant}$, correspond to Cauchy 
surfaces for de Sitter space.\footnote{  
A Cauchy surface is
any hypersurface such that every future-directed 
time-like vector intercepts it exactly once.
(See, {\it e.g.}, \cite{hawkellis}.)} 
Quantizing on a Cauchy surface in $C$ thus 
specifies all of the \lq\lq initial data" for the system,
including the initial conditions for the spatially-open universe
inside region $R$.

The equation of motion for a scalar field of mass $M$ in this space is
%%%%%%%%%%%%
\eq
\left(\Box - M^2 \right) \phi (x) = 
\left( (\sqrt{-g})^{-1} \partial_\mu \left[ (\sqrt{-g})
g^{\mu\nu} \partial_\nu \right]  - M^2 \right)
 \phi(x) = 0 ,
\label{c1eomC}
\en
and separation of variables gives a 
family of solutions $\phi_{k \ell m}(x)$.
To
quantize, $\phi$ and its canonically conjugate momentum 
$\Pi$ are promoted to Heisenberg 
operators, and expanded as
%%%%%%%%%%%%%%%%%
\eq
\hat{\phi} (x) = \sum_{k\ell m} \left[ \phi_{k \ell m} 
(x)
\hat{a}_{k \ell m} + \overline{\phi}_{k \ell m}(x) \hat{a}^\dagger_{k
\ell m}
\right] 
\label{c1hatfields}
\en
and similarly for $\hat{\Pi}$.
The creation and annihilation operators satisfy
$[\hat{a}_{k \ell m},\>
\hat{a}^\dagger_{k^\prime \ell^\prime m^\prime}] = \delta (k
-  k^\prime) \delta_{\ell^\prime \ell} \delta_{m^\prime m}$ with the
other commutators vanishing.
Here an overline denotes complex conjugation, $\sum_{k
\ell m}$ is a placeholder for the appropriate measure,
and the choice of vacuum state satisfying 
$\hat{a}_{k \ell m} \vert 0 \rangle
= 0$ for all $(k, \ell, m)$ provides a division into 
positive and negative frequency modes.
The fixed-time canonical commutation relations,
$[\hat{\phi}(x),\hat{\Pi}(x^\prime)] = i 
\delta^{3}(\vec{x} - \vec{x}^\prime)$ then imply that
%%%%%%%%%%%% 
\beqn
\nonumber (\phi_{k \ell m},\> \phi_{k^\prime \ell^\prime m^\prime}) 
&=& \delta
\left( k - k^\prime \right) 
\delta_{\ell^\prime \ell} \delta_{m^\prime m} 
\>\>, \>\> (\phi_{k \ell m},\>\overline{ \phi}_{
k^\prime \ell^\prime m^\prime} ) = 0\>\>,
\\
(\overline{\phi}_{ k \ell
m},\>\overline{ \phi}_{  k^\prime 
\ell^\prime m^\prime} ) &=& - \delta
\left( k - k^\prime \right) \delta_{\ell^\prime \ell} 
\delta_{m^\prime m}  ,  
\label{c1KG}
\eeqn
where the Klein-Gordon inner product is defined by
%%%%%%%%%%%%%%%%%
\eq
(\phi_{k \ell m},\>\phi_{k^\prime \ell^\prime m^\prime} ) \equiv -
\int_\Sigma \left(
\phi_{k \ell m}\>{\mathop{\partial_\mu}\limits^\leftrightarrow}\>
\overline{\phi}_{k^\prime \ell^\prime m^\prime} \right) \sqrt{-
g_\Sigma}
\> n^\mu\> d
\Sigma \; .
\label{c1KG2}
\en
Here $\Sigma$ is a (spacelike) Cauchy surface and $n^\mu$ is a
future-directed unit vector normal 
to this surface.  (The extra factor of
$i$ which multiplies the righthand side of 
equation (\ref{c1KG2}) in
\cite{birdavies} is absent here because of 
our different sign convention
for the metric.)  This inner product is independent of
Cauchy surface, and more generally is
independent of the choice of $\Sigma$, as long as the
fields fall off sufficiently quickly on the time-like boundaries.

The physically-motivated choice of initial 
vacuum state in 
models of inflation is the Bunch-Davies vacuum \cite{bunchdav},
which respects the symmetries of de Sitter space and
reduces to the Minkowski space vacuum 
at early times and over short distances.
For the flat slicing, the Bunch-Davies positive-frequency modes
are 
%%%%%%%%%%%%%%%
\eq
\phi_{k\ell m} (x_f) = \frac{k}{\sqrt{2}} \exp\left[ i \frac{\pi}{2}
\left( \nu + 
\frac{1}{2} \right) \right] 
\frac{ (-\eta)^{1/2}}{a (\eta)} H_\nu^{(1)}  (-k\eta) 
j_\ell (k r_f) Y_{\ell m} (\Omega) ,
\label{flatbasis1}
\en
where $\nu \equiv \sqrt{ {9 \over 4} - M^2}$,
$H_\nu^{(1)} (z)$ 
is a Hankel
function of the first kind, $j_{\ell} (z)$ is a spherical 
Bessel function,
and $Y_{\ell m} (\Omega)$ is the usual spherical harmonic.  
The measure in equation (\ref{c1hatfields}) for the expansion 
in flat modes $\phi_{k \ell m}$ is $\sum_{k \ell m} = 
\int_0^\infty d 
k \sum_{\ell = 0}^\infty \sum_{m = - \ell}^\ell$.  
%jjj took out next sentence, seemed redundant?  jjj

For the hyperbolic slicing with the metric $ds_c^2$ of equation
(\ref{c1openmetrics}), the positive-frequency
solutions to the equations of motion are \cite{bgt,sasaki95,buchtur95}
\eq
u_{p \ell m}(x) = \frac{\chi_p (t_c)}{a(t_c)} f_{pl}(r_c) Y_{\ell
m}(\Omega) ,
\en
where
\beqn
\nonumber \chi_p(t_c) &=&  \alpha_p(\nu^\prime) P^{-
ip}_{\nu^\prime}
(\sin t_c) + 
\beta_p(\nu^\prime) P^{ip}_{\nu^\prime} (\sin t_c) \; ,\\
f_{p \ell} (r_c) &=& \frac{P^{-\ell - 1/2}_{ip - 1/2} (i \sinh
r_c)}{\sqrt{i\cosh r_c}} , 
\label{chip1}
\eeqn
and
\eq
\nu^\prime \equiv \nu - {1 \over 2} = \sqrt{{9 \over 4} - M^2} - 
{1 \over 2} 
\; .
\en
Here $P^\mu_\nu (z)$ is an associated Legendre function of the first
kind, while
the specific forms of $\alpha_{p}, \beta_p$ will not be needed.
Within region $C$, the 
$f_{p \ell}
(r_c)$ play the role of positive-frequency solutions, and the
$\chi_p (t_c)$ are spatial eigenfunctions; when continued into region
$R$, these roles are reversed. 

%In region $C$, the equation of motion (\ref{c1eomC}) takes the form,
%\eq
%\left[ \frac{1}{a^3 (t_c)} \frac{\partial}{\partial t_c} \left(
% a^3 (t_c)
%\frac{\partial}{\partial t_c} \right) - \frac{1}{a^2 (t_c)} 
%{\bf L}^2 -
%M^2 \right] \phi (t_c,\>r_c,\>\Omega) = 0 ,
%\en
%where the operator ${\bf L}^2$ (which is 
%the spatial Laplacian in region $R$) has the eigenvalues
%\eq
%{\bf L}^2 (f_{p \ell}(r_c) Y_{\ell m}(\Omega)) = - (1 + p^2) 
%(f_{p \ell}(r_c)
%Y_{\ell m}(\Omega)) .
%\en  jjj took this out, don't know, is it ok? jjj
A Friedmann-Robertson-Walker 
metric with spatial curvature $K$ and 
cosmic scale factor $a(t)$ has a
physical curvature length scale $a(t)/\vert K \vert$.
For a flat universe the comoving curvature length scale thus
runs off to infinity, whereas
in a spatially closed or open
metric, the comoving curvature length scale
is $+1$.  Eigenvalues of the (region $R$) spatial Laplacian in this background 
are $-(k/a)^2$, where $(k/a)$ is the
inverse of a physical length:  $k/a = 1/x_{\rm phys}$, with $0
\leq k^2 < \infty$. 
Defining $p^2 \equiv k^2 - 1$, $p^2 > 0$ for
subcurvature modes, and $-1 < p^2 \leq 0$ 
for supercurvature modes.
The continuum of subcurvature modes $0 \le p \le \infty$ 
is sufficient
to describe a Gaussian random field in region $R$, 
see \cite{lythwos} for detailed discussion.
In addition, for $p^2 < 0$, 
inner products of the form (\ref{c1KG2}) on fixed-time 
(non-Cauchy)
surfaces within region $R$ diverge, and so all supercurvature 
modes naively appear to be unnormalizable.
Studying quantization and completeness more appropriately
on a fixed time $r_c$ Cauchy surface in region $C$,
it was found in \cite{sasaki95} that for $M^2 < 2 H^2$ (restoring the
Hubble radius, $H^{-1}$), supercurvature
modes are normalizable in vacuum for a discrete
value of $p$.
In addition, it was shown there that this value of $p$ must be
included to obtain the correct Wightman function for the 
Bunch-Davies vacuum.  (See also \cite{Mosch}.)

This discrete normalizable supercurvature mode
can be understood as follows \cite{sasaki95,buchtur95}.
Its presence is suggested by
an analogy between these supercurvature
modes and bound states in a potential.
The only dependence on $M$ in the wavefunctions is in the
\lq spatial' (in region $C$)
eigenfunctions $\chi_p(t_c)$.
Defining $\sin t_c \equiv \tanh u$,
the equation of motion for $\chi_p$ becomes
%%%%%%%%%%%%%%%%%%
\beqn
\nonumber \left[ - 
\frac{d^2}{d u^2} + U(u) \right] \chi_p &=& p^2
\chi_p \; , \\
U(u) &\equiv& \frac{M^2 - 2}{\cosh^2 u} .
\eeqn
This is a one-dimensional Schr\"{o}dinger-like equation with
the potential $U(u)$ and energy $p^2$.  As noted in
\cite{garriga,sasaki96,buchtur95}, the potential $U(u)$ vanishes as
$u \rightarrow
\pm \infty$, revealing that in this limit there exists a continuous
spectrum of modes with $p^2 \geq 0$.  But over finite intervals of
$u$, if the mass of the field satisfies $M^2 < 2$, the potential
$U(u)$ has a valley and the modes $\chi_p$ behave as discrete bound
states with $p^2 < 0$.  
As DeWitt has shown, such discrete states in a
field's spectrum are generic for fields quantized 
on compact subspaces. \cite{dewitt} 

The inner product on this space is proportional to 
\eq
\int_{-\infty}^\infty du \> 
\chi_p(\tanh u) \overline{\chi}_{p^\prime} (\tanh u) \; ,
\en
so for normalizability, $\chi_p(\tanh u)$ must be
bounded as $u \to \pm \infty$.
(A similar argument is found in the appendix of \cite{sasaki95}.)
A supercurvature mode has
$p^2 <0$ and $p^2 +1 \ge 0$, so define $p = i \Lambda$,
with $0 \leq \Lambda \leq 1$, and $\Lambda$ real.
The asymptotics of $\chi_p$ near
$\tanh u \rightarrow \pm 1$ yields
(cf. \cite{buchtur95}):
%%%%%%%%%%%%%%%%
\eq
\Gamma (1 \mp ip) P^{\pm ip}_{\nu^\prime} (\tanh u) \sim e^{\pm
ipu}\>\>,\>\> u \rightarrow \infty .
\label{c1limit1}
\en
This is finite only for $P^{+ip}_{\nu^\prime}$ 
since $ip = - \Lambda < 0$,
thus $\beta_p = 0$ for the supercurvature modes.
The limit of $P^{ip}_{\nu^\prime}(\tanh u)$ as $u \to -\infty$
%can be found via several hypergeometric function identities
%(see \cite{Absteg}, equations 15.3.3 and 15.3.6), as done in 
is \cite{buchtur95}\footnote{Note that in the derivation 
of this result in \cite{buchtur95}, their equation (2.9)
is incorrect, including only the factor $w^{-\mu}$ instead of
$[(1-w)/w]^\mu$, though their next equation is correct.}
\eq
\Gamma(1 + \Lambda) P^{- \Lambda}_{\nu^\prime} (\tanh u) \sim
\frac{\Gamma(1 + 
\Lambda)\Gamma(\Lambda)}{\Gamma(1+{\nu^\prime} 
+
\Lambda)\Gamma( \Lambda - {\nu^\prime})}e^{ \Lambda \vert u 
\vert} +
\frac{\Gamma(1 + \Lambda)\Gamma(- \Lambda)}{\Gamma
(-{\nu^\prime})\Gamma(1+{\nu^\prime})}e^{- \Lambda \vert u 
\vert}\>,\> u
\rightarrow - \infty .
\en
As $0 < \Lambda \le 1$ is non-negative and $ 0 \le {\rm Re} \;
 {\nu^\prime} \le 1$,
the coefficient of $e^{\Lambda |u|}$ vanishes
only if $\Lambda = {\nu^\prime}$, producing an isolated
value of $p$ which is normalizable.
(Note that for
subcurvature modes, with $p$ real and non-negative, 
the solution $\chi_p
(t_c)$ simply oscillates at both endpoints and so remains
finite.)

The Klein Gordon normalized supercurvature modes are then 
\beqn
\nonumber
  u_{\Lambda \ell m} (x_c) &=& N_{\Lambda \ell m} (-
i\cos
t_c)^{{\nu^\prime} -1} \frac{P^{-\ell - 1/2}_{-\Lambda 
- 1/2} (i\sinh r_c
)}{\sqrt{i\cosh r_c}} Y_{\ell m} (\Omega) , \\
N_{\Lambda \ell m} 
&\equiv& \left[ \frac{\Gamma ({\nu^\prime} + 1/2) 
\Gamma  (-{\nu^\prime} +
\ell + 1) \Gamma ({\nu^\prime} + \ell + 1)}{2 \sqrt{\pi}
 \Gamma({\nu^\prime})} \right]^{1/2}
 \; .
\label{c1phi-Lambda}
\eeqn
%%%%
The scalar field in region $C$ is expanded as \cite{sasaki95}
%%%%%%%%%%%%%%%%%%%%
\beqn
\nonumber \hat{\phi}(x_c) &=& \int_0^\infty dp 
\sum_{\ell, m} \left[
u_{p \ell m} (x_c) \hat{a}_{p\ell m} + {\rm H.c.} \right] +
\sum_{\ell,m} \left[ u_{\Lambda \ell m} (x_c) 
\hat{a}_{\Lambda \ell  m}
+ {\rm H.c.} \right] \\
&\equiv& \hat{\phi}_{(p)} (x_c) + \hat{\phi}_{(\Lambda)} (x_c),
\label{c1phiopen}
\eeqn
where \lq\lq H.c" denotes the Hermitian conjugate, 
and $\sum_{\ell,m}
\equiv \sum_{\ell = 0}^\infty \sum_{m=-\ell}^\ell$.
Once quantized, these modes may be continued into region $R$. 
In the presence of a bubble wall \cite{sasaki96}, rather than 
in the vacuum,
the value of $\Lambda$ may change and 
a supercurvature mode may appear even
for $M^2 > 2$,
depending on the details of the model.  The normalizability
conditions may be solved for numerically.  In the absence of 
gravity,
there is also a supercurvature mode for the fluctuations of the 
bubble wall itself, which appears to become singular once
gravity is included
\cite{wallb}. 

\section{Overlap of Supercurvature with Flat Modes for a Massless Field}

We now re-express these supercurvature modes $\uel$ in terms of the
spatially-flat modes 
$\phi_{k \ell m}$ for a massless scalar field.  There are two things to
consider, however, before proceeding to the calculation.  First,
for $M = 0$, the $\ell = 0$ supercurvature mode of equation
(\ref{c1phi-Lambda}) is
constant.  Consequently its Klein-Gordon inner product is zero and its
norm (proportional to the square root of the inverse of this norm)
diverges as $\Gamma^{1/2} (\ell = 0)$.  Including this state naively in
the Wightman function sum over states will thus diverge as well.  It is
known that the Wightman function for a massless, minimally-coupled 
scalar
field in de Sitter spacetime is infrared divergent \cite{desinfra,kg93},
and in this
particular slicing, the specific state $u_{\Lambda = 1, 00}$ appears to
be
the lone source of the divergence. \\
\indent 
The $r_c$-dependent portion 
of the equations of motion in region $C$, are exactly like a
$(2,1)$-dimensional spacetime with scale factor $a(r_c) = \cosh r_c$,  
For positive frequency modes,
%%%%%%%%%
\eq
\left[ \frac{1}{a^2 (r_c)} \frac{\partial}{\partial r_c} 
\left( a^2 (r_c) 
\frac{\partial}{\partial r_c} \right) + 
\left( 1 + p^2 \right) + \frac{ \ell (\ell 
+ 1 )}{a^2 (r_c)} \right] \{f_{pl} (r_c) Y_{\ell m} 
(\Omega) \} = 0 \; 
\en
and the \lq frequency' $\omega_{r_c}$ associated with the 
time-coordinate may then be written
%%%%%%%%%%%
\eq
\omega_{r_c}^2 = (1 + p^2 ) + \frac{ \ell (\ell + 1)}{a^2 (r_c)} ,
\en
the analogue of $\omega^2 = M^2 + k^2 / a^2 (t)$. 
This $\omega_{r_c}$
vanishes only for $p = i$ and $\ell = 0$, indicating a
zero mode corresponding to the symmetry $\phi \rightarrow \phi +
{\rm constant}$.
This zero mode should be replaced by a collective 
coordinate.\footnote{Such divergences do not directly affect CMB
anisotropy calculations, which correspond to $\ell > 0$.} 
\cite{kg93,jdcdk,rajaraman}  Zero modes have also been identified in
the context of two field models in \cite{gar-97}.  
In the following, we address only 
the finite modes $(\ell > 0)$ for the massless, minimally-coupled case.

The only Cauchy surface for the entire de Sitter spacetime in region 
$F$ is the limiting curve $\eta \to 0$, corresponding to
$t_r \to \infty$.  This is in contrast with
region $C$ where any `time'
$r_c = {\rm constant}$ surface is a Cauchy surface. 
However, all of region $R$,
containing our open observable universe in models of 
open inflation, 
is contained within region $F$.  Because our aim is to provide
a useful heuristic relation between the unusual supercurvature modes and
the more familiar spatially-flat modes, we will thus work in this section
at fixed time within region $F$.  Verifying these results,
Appendix B contains
a parallel calculation for all odd $\ell$ and $\ell = 2$ along the proper
Cauchy surface $r_c = 0$.  In
Appendix C the overlap for $\ell = 1$ is found 
on the boundary of $F$, corresponding to 
Cauchy surface $\eta = 0$.

The Klein-Gordon inner product in region $F$ is
%%%%%%%%%%%%%%%
\eq
(u, \> v) = - i a^2 (\eta) \int_0^\infty dr_f \> r_f^2 \int d \Omega
\left[ u \left( \partial_\eta \overline{v} \right) - 
\left( \partial_\eta
u \right) \overline{v} \right]_\Sigma .
\en
This can be evaluated along any fixed-$\eta$ surface $\Sigma$, and, if
the fields fall off sufficiently quickly on the time-like boundaries,
will
be independent of the specific choice of $\eta$, even though such
fixed-$\eta$ surfaces are not Cauchy surfaces for the entire spacetime.
The modes $\phi_{k \ell m} (x_f)$ and $\overline{\phi}_{k \ell m}
(x_f)$ satisfy the inner product relations of equation (\ref{c1KG})
along such fixed-$\eta$ surfaces within region $F$, and
form a complete
set of orthonormalized modes within region $F$; thus they may be
used to expand any normalizable function within region $F$.
The massless supercurvature mode
$u_{\Lambda=1,\ell^\prime m^\prime}$ is normalizable
on fixed $\eta$ surfaces with $-1 \le \eta \le 0$, as
shown in Appendix A.  Its norm being independent of $\eta$
suggests that its falloff is fast enough to make the inner
products independent of $\eta$ as well.
Thus we expect we can
express the massless supercurvature mode $u_{\Lambda = 1,
\ell^\prime m^\prime} (x_r (x_f))$ in terms of the flat basis 
functions as 
\eq
u_{\Lambda = 1, \ell^\prime m^\prime} (x_r(x_f)) = \int_0^\infty dk
\sum_{\ell, m} \left[ \alpha_{k \ell m} \> \phi_{k \ell m} (x_f) +
\beta_{k \ell m} \> \overline{\phi}_{k \ell m} (x_f) \right] ,
\label{ualpha1}
\en
with $\alpha_{k \ell m}$ and $\beta_{k \ell m}$ constant complex
coefficients,
%%%%%%%%%%%%%%%%%%%%
\eq
\alpha_{k\ell m} = \left( u_{\Lambda = 1, \ell m}, \> \phi_{k \ell m}
\right) \>\>,\>\> 
\beta_{k \ell m} = - \left( u_{\Lambda = 1, \ell m}, \>
\overline{\phi}_{k \ell m} \right) ,
\label{alphabeta}
\en
even for a fixed $\eta$ (non-Cauchy) surface. 

For $M = 0$, the spatially-flat basis modes (equation 
(\ref{flatbasis1})) reduce to
\beqn
\nonumber \phi_{k \ell m} (x_f) &=& 
\frac{i}{\sqrt{\pi k}} (1 + i k \eta)
e^{-i k \eta} j_\ell (kr_f) Y_{\ell m} (\Omega) \\
&\equiv& \frac{i}{\sqrt{\pi k}} F_{k \ell } (\eta, r_f) Y_{\ell m}
(\Omega),
\label{flatF}
\eeqn
and the normalized supercurvature modes for the massless case are
%%%%%%%%%%%%%%
\beqn
\nonumber u_{\Lambda = 1, \ell m} (x_r) &=& \frac{1}{2} \sqrt{\Gamma
(\ell) \Gamma (\ell + 2)} \> \frac{P^{-\ell - 1/2}_{-3/2} (\cosh
r_r)}{\sqrt{\sinh r_r}} \> Y_{\ell m} (\Omega) \\
&\equiv& \frac{1}{2} \sqrt{\Gamma (\ell) \Gamma (\ell + 2)} \> S_\ell
(r_r) \> Y_{\ell m} (\Omega) 
\label{uR}
\eeqn
within region $R$.
Using the embedding coordinates $z^\mu$ of the five-dimensional
Minkowski space to relate the coordinates $x_f$ and $x_r$ (see equations
(\ref{c1xflat2}) and (\ref{c1xopen})) yields 
%%%%%%%%%%%%%%%%%%%
\eq
\cosh r_r = \frac{g}{\sqrt{g^2-1}} \>\>,\>\> \sinh r_r =
\frac{1}{\sqrt{g^2-1}} \; ,
\label{rtoflat}
\en
where
\eq
g = g (\eta, r_f) \equiv \frac{1}{2 r_f} \left( 1 + r_f^2 - \eta^2
\right) .
\label{g}
\en

The coordinate $g$ is convenient because
of the identities (see \cite{Absteg}, equations 8.2.7 and
8.6.7):
%%%%%%%%%%%%%%%%%%
\beqn
\nonumber P^{-\alpha - 1/2}_{-\beta - 1/2} \left( \frac{z}{\sqrt{z^2-1}}
\right) &=& \sqrt{\frac{2}{\pi}} \frac{ (z^2-1)^{1/4} \> e^{-i \beta \pi}
\> Q^\beta_\alpha (z)}{\Gamma (\alpha + \beta + 1 )} , \\
Q_\ell^1 (z) &=& \sqrt{z^2 - 1} \> \frac{dQ_\ell (z)}{dz} ,
\label{PQidentities}
\eeqn
where
$Re\> [z] > 0$.  Taking $g = z$ and noting that 
$0 \le r_f \le \infty$ in the inner product means that
we want fixed $\eta$ surfaces with $ -1 \le \eta \le 0$.
When $z^2 = g^2 < 1$, the argument of
$P^{-\alpha-1/2}_{-\beta-1/2}$ becomes complex and hence the righthand
side should be understood with $z$ having a small imaginary part.
Unless $\eta = 0$, fixing $\eta$ and letting $r_f$
range over its values $0 \le r_f \le \infty$ will include some values
of $g^2 < 1$.

With the identities (\ref{PQidentities}) above,
%%%%%%%%%%%%%%%%%%%%%
\eq
\nonumber S_\ell (r_r) = \frac{P^{-\ell - 1/2}_{-3/2} (\cosh
r_r)}{\sqrt{\sinh r_r}} 
= - \sqrt{\frac{2}{\pi}} \frac{1}{\Gamma (\ell + 2)} \left( g^2 - 1
\right) \>\partial_g Q_\ell (g) ,
\en
and the overlap in equation (\ref{alphabeta}) is then
%%%%%%%%%%%%%%%
\beqn
\nonumber \alpha_{k \ell m} &=& - {\cal C}_{k \ell} (-\eta)^{-2}
e^{ik\eta} \int_0^\infty dr_f \> r_f^2 \> j_\ell (k r_f) \left[ k^2 \eta 
\> S_\ell (r_r(x_f)) - (1-ik\eta) \partial_\eta S_\ell (r_r(x_f)) 
\right]_\Sigma , \\
{\cal C}_{k \ell} &\equiv& \frac{1}{2} \left[ \frac{\Gamma (\ell) \Gamma
(\ell + 2)}{\pi k} \right]^{1/2} .
\label{alpha1}
\eeqn
Note that the constant coefficients ${\cal C}_{k \ell}$ are purely real.
Similarly,
%%%%%%%%%%%%%%%%%
\eq
\beta_{k \ell m} = {\cal C}_{k\ell} (-\eta)^{-2} e^{-i k \eta} (-1)^m
\delta_{m, - m^\prime} \left[ k^2 \eta - (1 + ik \eta) \partial_\eta
\right] \int_0^\infty dr_f \left[ r_f^2 \> j_\ell (k r_f) S_\ell (r_r)
\right]_\Sigma .
\en
As both $\phi(x_f), u(x_r)$ correspond to positive frequency for the
Bunch-Davies vacuum, 
$\beta_{k \ell m} = 0$ for all $k$,
$\ell$, and $m$, giving
%%%%%%%%%%%%%%%%%
\beqn
\nonumber \partial_\eta I_\ell (k)_{\vert \Sigma} &=& 
\frac{k^2 \eta}{(1 +
i k \eta)} \> I_\ell (k)_{\vert \Sigma} , \\
I_\ell (k)_{\vert \Sigma} &\equiv& \int_0^\infty dr_f \left[ r_f^2 \>
j_\ell (k r_f) S_\ell (r_r) \right]_\Sigma , 
\label{I_ell}
\eeqn
or
%%%%%%%%%%%%%
\eq
\alpha_{k \ell m} = - 2 i {\cal C}_{k \ell} k^3 
\frac{e^{ik\eta}}{(1 + i k
\eta)} I_\ell (k)_{\vert \Sigma} .
\label{alpha2}
\en

Equation (\ref{I_ell}) may be used to demonstrate explicitly that
$\partial_\eta \alpha_{k \ell m} = 0$ identically, allowing a
choice of any convenient value of $\eta$ to evaluate $I_\ell (k)$.
We choose $\eta = -1$ in this section; the case
$\eta = 0$ and $\ell = 1$ is in Appendix C. 

Along the surface $\eta = -1$, $g_* = r_f/2$ and
%%%%%%%%%%%%%
\eq
I_\ell (k)_{\vert \Sigma} = -8 
\sqrt{\frac{2}{\pi}} \frac{1}{\Gamma (\ell
+ 2)} \int_0^\infty dg_* \> g_*^2  \left( g_*^2 - 1 \right)
 \> j_\ell (2 k
g_*) \frac{d Q_\ell (g_*)}{d g_*} ,
\label{I_ell2}
\en
where the $*$ 
indicates that a particular value of $\eta$ has been chosen
for this evaluation.  For
several small values of $\ell$,
repeated integration by
parts gives the general form (dropping the
subscript $*$ )
%%%%%%%%%%%%%%%%
\beqn
\nonumber \int dg \> g^2 \left( g^2 - 1 \right) j_\ell (2 k g)
\partial_{g} Q_\ell (g) &=& a_1\> \cos (2kg) + a_2\> \sin (2kg) \\
\nonumber &+& a_3 \left( {\rm ci} \> [2k(1+g)]  - {\rm ci} \> [2k(1-g)]
\right) \\ 
\nonumber &+& a_4 \left( {\rm Si}\> [ 2k (1+g) ] - {\rm Si} \>[2k(1-g)]
\right) \\
 &+& a_5 \> {\rm Si}\>[2kg] ,
\eeqn
where
%%%%%%%%%%%%%%%
${\rm ci}\>(z) \equiv  - \int_z^\infty dt \> \frac{\cos t}{t}$,
${\rm Si} \> (z) \equiv \int_0^z dt \> \frac{\sin t}{t}$,
and generally $a_i= a_i(g,k)$.
For $\ell = 1$, the nonzero coefficients are
%%%%%%%%%%%%%%%%%%
\beqn
\nonumber a_1 &=& \frac{g}{2k^3} + \frac{1}{8k^5} \ln \left(
\frac{g+1}{g-1} \right) \left[ 1 + k^2 - 2k^2g^2 \right] , \\
\nonumber a_2 &=& \frac{(k^2 g^2 - 2)}{4k^4} + \frac{1}{8k^4} \ln \left(
\frac{g+1}{g-1} \right) \left[ 2g + k^2 g \left( 1 - g^2 \right) \right] ,
\\
\nonumber a_3 &=& \frac{1}{8k^5} \left[ (k^2 - 1) \> \cos (2k) - 2k \> \sin
(2k) \right] , \\
a_4 &=& \frac{1}{8k^5} \left[ 2k \> \cos (2k) + (k^2 - 1) \> \sin
(2k) \right] .
\label{ai1}
\eeqn
Using \cite{Absteg,gr}, as $g \rightarrow
\infty$,
$a_1 \rightarrow 0$ and $a_2 \rightarrow (6k^2)^{-1}$ and
%.  From \cite{gr},
%equation
%8.233.2 and \cite{Absteg}, equations 4.1.14, 5.2.19, and 5.2.25, we also
%have
%%%%%%%%%%%%%%%
%\beqn
%\nonumber {\rm ci} \> (z) - {\rm ci} \> (z \> e^{\pm i \pi} ) &=& \mp i
%\pi , \\
%\nonumber {\rm Si}\> (z) - {\rm Si}\> (-z) &=& 2 \> {\rm Si} \> (z) , \\
%\lim_{z \rightarrow \infty} \> {\rm Si} \> (z) &=&
%\frac{\pi}{2} , 
%\eeqn
%yielding for the definite integral
%%%%%%%%%%%%%%%
thus
\eq
\int_0^\infty dg \> g^2 \left( g^2 - 1 \right) 
j_1 (2kg) \partial_g Q_1
(g) = \frac{1}{6k^2} \sin (2kg)_{g \rightarrow \infty}
 + \frac{\pi (1 -
ik) e^{ik}}{4k^5} \left[ k \> \cos (k) - \sin (k) \right] .
\en
Based on comparison with the explicit calculation of this 
definite
integral along different surfaces (see Appendix B), we drop
the
first term, as its limiting value
oscillates at $g \rightarrow \infty$. 
Then
the coefficient of expansion
$\alpha_{k, \ell = 1, m}$ becomes
%%%%%%%%%%%%%%%%%
\eq
\alpha_{k, \ell = 1, m} = -2 i\> k^{-1/2} \> j_1 (k) .
\label{alphaone}
\en
Repeating the same analysis for 
$\ell = 2,3$, for which
$a_1$ and $a_2$ both vanish
identically as $g \rightarrow \infty$, yields
%%%%%%%%%%%%%
\beqn
\nonumber  \alpha_{k, \ell = 2 , m} &=& -2 i \sqrt{3}
  \> k^{-1/2}\> j_2 (k) , \\
\alpha_{k , \ell = 3 , m} &=& - 2 i \sqrt{6}  
\> k^{-1/2} \> j_3 (k) .
\label{alphas}
\eeqn
These first three expansion coefficients are plotted in Figure 3.
%%%%%%%%%%%%%%%%%%%%%%
\begin{figure}
\centerline{\epsfig{file=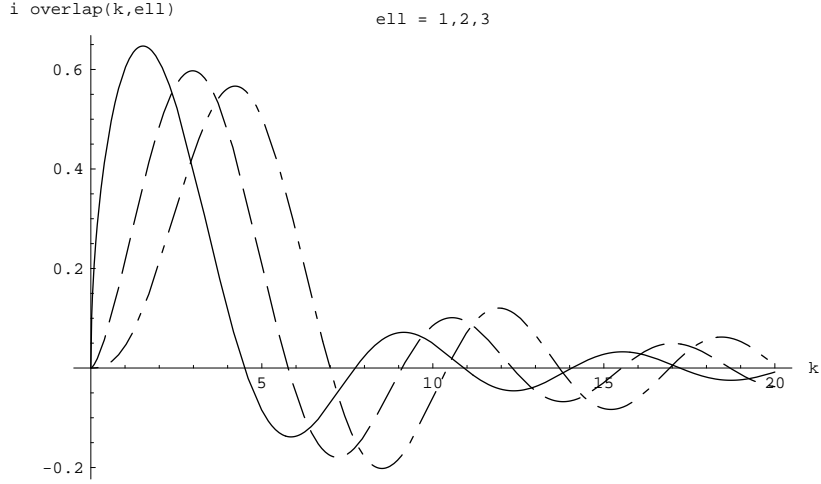,height=3.0in}}
\vspace{10pt}
\caption{\small The first three expansion coefficients, 
$i\alpha_{k \ell
m}$ for $\ell = 1,\>2,\>3$, plotted against wavenumber $k$ of
the modes in the spatially-flat slicing, $x_f$.}
\end{figure}
These coefficients are finite in both the $k \rightarrow 0$ and $k
\rightarrow \infty$ limits,
%%%%%%%%%%%%%
\beqn
{\displaystyle
\lim_{k\rightarrow 0}} \> k^{-1/2} \> j_\ell (k) & \sim &
{\rm constant} \> k^{\ell - 1/2}  \to 0 \; \;\> ({\rm for}
 \>\> \ell > 0)\\
{\displaystyle
\lim_{k \rightarrow \infty}} \> k^{-1/2} \> j_\ell (k) &\sim&
{\rm constant} \> k^{-3/2} \> 
\sin \left( k - {\pi \ell \over 2} \right) \to 0
\eeqn
by the asymptotics of spherical
Bessel functions ({\it e.g.} \cite{Absteg}, equations
10.1.4, and 9.2.5).

Equations (\ref{alphaone},\ref{alphas})
suggest that the general form for the overlap of the supercurvature
modes with the flat basis functions is proportional to $j_\ell(k)
k^{-1/2}$.
This can be tested by seeing if these postulated $\alpha_{k \ell m}$
reconstruct the supercurvature mode, {\it i.e.}
%%%%%%%%%%%%%%%%
\eq
u_{\Lambda = 1, \ell m} (x_r) = \int_0^\infty 
dk \> \alpha_{k \ell m} \> \phi_{k \ell m} \; . 
\label{ovdef}
\en

The most convenient choice of $ \eta$ for this integral is the surface
$\eta = 0$, since in this case the $k$ dependence in
$\phi_{ \ell m}$ is proportional to $ j_\ell(k r_f) k^{-1/2}$.
Writing $\alpha_{k \ell m} = {\cal D}_\ell \> k^{-1/2} \> j_\ell
(k)$ and substituting into equation (\ref{ovdef}), we have 
%%%%%%%%%%%%%%%%%
\eq
\frac{1}{2} \sqrt{\Gamma (\ell) \Gamma (\ell + 2)} \> S_\ell (r_r(\eta = 0)) 
=^{?} 
\frac{i}{\sqrt{\pi}} \> {\cal D}_\ell \int_0^\infty dk \> \frac{j_\ell
(kr_f) j_\ell (k)}{k} .
\label{alphacheck}
\en
For the left hand side of this equation, 
using equation (\ref{g}) and setting $\eta = 0$,
\eq
S_\ell (r_r(\eta = 0)) = 
\sqrt{ \frac{1 - r_f^2}{2 r_f}} \> P^{-\ell - 1/2}_{-3/2}
\left( \frac{1 + r_f^2}{1 - r_f^2} \right) .
\en
Region $R$ corresponds to $z^4 \geq 1$, which requires 
$r_f^2 \le 1$ as $\eta \to 0$.

The right hand side of equation (\ref{alphacheck}) can be integrated to give
(\cite{Absteg}, equation 11.4.34):
%%%%%%%%%%%%%%%%%%
\eq
\int_0^\infty dk \> \frac{j_\ell (kr_f) j_\ell (k)}{k} =
\frac{\sqrt{\pi}}{4} \frac{r_f^\ell \> \Gamma (\ell)}{\Gamma (\ell + 3 /
2 )} \> F \left( \ell , - 1/2 ; \ell + 3/2; r_f^2 \right) ,
\label{jint}
\en
where $F (a,b;c;z)$ is the hypergeometric function.  By using
$F (a,b; c;z) = F(b,a; c; z)$, the representation of
$P^\mu_\nu (z)$ in terms of hypergeometric functions
(\cite{gr}, equation 8.772.3), and
$P^\mu_\nu (z) = P^\mu_{-\nu - 1} (z)$,
the righthand side of equation (\ref{alphacheck}) becomes
\beqn
\nonumber \frac{\sqrt{\pi}}{4} 
\frac{r_f^\ell \> \Gamma (\ell)}{\Gamma (\ell + 3 /
2 )} \Gamma(\ell + 3/2)\>r_f^{-\ell - 1/2}\sqrt{1-r_f^2}
\> P^{-\ell - 1/2}_{-3/2}
\left(\frac{1+r_f^2}{1-r_f^2}\right) = \\
\frac{\sqrt{\pi}}{4} \Gamma(\ell) \sqrt{\frac{1 - r_f^2}{r_f}}
\> P^{-\ell - 1/2}_{-3/2}
\left(\frac{1+r_f^2}{1-r_f^2}\right) .
\eeqn

Thus the $r_f$ dependence on both sides of
equation (\ref{alphacheck}) matches exactly, and the constant 
${\cal D}_\ell$ may be read off:
%%%%%%%%%%%%%%%%%%
\eq
{\cal D}_\ell = -i \sqrt{2 \ell (\ell + 1)} .
\en
This coefficient
reproduces the specific  $\alpha_{k \ell m}$ calculated earlier
for $\ell = 1, \> 2, \> 3$ along the surface $\eta = -1$.  
Resumming the equation for $r_f >1$ would correspond to region
$L$ in de Sitter space.

Thus we
conclude that the constant coefficients $\alpha_{k \ell m}$ which
relate the normalized supercurvature modes $u_{\Lambda \ell m}$ and the
spatially-flat basis modes $\phi_{k \ell m}$ are:
%%%%%%%%%%%%%%%
\eq
\alpha_{k \ell m} = 
-i \sqrt{2 \ell ( \ell + 1 ) } \> k^{-1/2} \> j_\ell(k). 
\label{result}
\en
This is the main result of this paper.

Taking the limit $r_f \rightarrow 1$ in equation (\ref{jint}), 
and using the identity (\cite{gr}, equation 9.122.1):
\eq
F(\alpha,\beta,\gamma;1) = \frac{\Gamma(\gamma) \Gamma(\gamma-\alpha-\beta)}
{\Gamma(\gamma-\alpha)\Gamma(\gamma - \beta)} \; ,
\en
it is easy to verify in addition that
\eq
\int_0^\infty dk \> \vert \alpha_{k \ell m} \vert^2 = 2 \ell
(\ell + 1) \int_0^\infty dk \> \frac{ j_\ell (k) j_\ell (k)}{k} = 1 .
\en

\section{Conclusion}
In conclusion, we have given the explicit form for the overlap between
the flat basis functions (\ref{flatF}) and
the massless supercurvature modes.  As a result, the
supercurvature modes within that patch of de Sitter space which would
contain our open observable universe can be written
\eq
\uel (x_r(x_f)) = -i \sqrt{2\ell (\ell + 1)} \int_0^\infty dk \> k^{-1/2} 
\> j_\ell (k) \> \phi_{k\ell m} (x_f) .
\en
The long-wavelength supercurvature modes are 
distributed over the spatially-flat basis modes, oscillating over
flat-space comoving wavenumber $k$ with decreasing amplitude.
More quantitatively,
the spherical bessel functions $j_\ell(k)$ have their first and
largest maximum (\cite{Absteg} 10.1.59)
near $k \sim (\ell + \frac{1}{2}) [1 + O(\ell^{- 2/3})]$
with the approximation improving as $\ell$ increases, but
the damping envelope going as 
$k^{-1/2}$ lowers this peak for higher $\ell$.
It may be possible to use the description \cite{gar-97} of
$M^2 \ne 0$ supercurvature modes 
as small perturbations of the massless supercurvature
modes to extend the above to small $M^2$.

This expression for the supercurvature mode on fixed $\eta$ surfaces,
extending into region $R$, may be useful for understanding
the effects of supercurvature modes during reheating.
Unlike the event of bubble nucleation, reheating occurs
in the future of region $C$ and hence descriptions using 
fixed time $r_c$ surfaces are 
not appropriate.  Rather, it is important to understand the dynamics of
these modes within region $R$, corresponding to our observable, open
universe.  Having an expression for the normalizable 
supercurvature modes on slicings extending into region $R$ is a step 
in separating long wavelength properties of these modes from issues 
related to their non-normalizability at fixed time $t_r$ in region $R$.

We showed by comparing Cauchy and non-Cauchy surface
calculations that in some cases non-Cauchy surface calculations of norms
(in the appendix) and overlaps (in the text) agree for supercurvature
modes, up to an identifiable boundary term. These non-Cauchy surfaces
(fixed time in the flat coordinates)  extend into the open universe and
thus could be used (with caution) to calculate other properties.
The complementary calculations along different surfaces were required
to verify that the non-Cauchy surface representation was indeed correct.

In addition, we identify one specific supercurvature mode as
responsible for the well-known infrared divergence for massless scalar
fields in de Sitter space.  Not only does this mode have divergent norm;
it is demonstrated here to be a dynamical zero-mode.  This identification
is a necessary first step toward its eventual replacement with an
appropriate collective coordinate, similar to what has been done in closed
coordinates\cite{kg93}.  Something similar has been done in \cite{gar-97}
in the context of two field models and quasi-open
inflation, here it is found
more generally as a property of the massless supercurvature modes.

\section{Appendices}
\subsection{Normalization of supercurvature mode at 
fixed time $-1 \le \eta \le 0$}
In region $F$ the supercurvature mode can be written as
(using equation \ref{rtoflat})
\eq
 S_\ell (r_r(g)) = \frac{P^{-\ell - 1/2}_{-3/2} (\cosh
r_r)}{\sqrt{\sinh r_r}} =
(g^2 - 1)^{1/4} \> P^{-\ell - 1/2}_{-3/2}\left(\frac{g}{\sqrt{g^2
-1}}\right) 
\en
which can be used to extend out of region $R$, to where $g \le 1$.
We will consider only $ -1 \le \eta \le 0$ for convenience.

The inner product for fixed time $\eta$ is
\eq
(S_\ell, \> S_\ell) = - i a^2 (\eta) \int_0^\infty dr_f \> r_f^2 \int d
\Omega
\left[ S_\ell \left( \partial_\eta \overline{S}_\ell \right) - \left(
\partial_\eta
S_\ell \right) \overline{S}_\ell \right]_\Sigma .
\en
The integral over $\Omega$ gives a delta function and will be
suppressed.

For $g \ge 1$, $S_\ell = \overline{S}_\ell$ because both the associated 
Legendre function
and its argument are real, and so the integrand disappears.
For fixed $\eta$, $g^2 = 1$ corresponds to
\eq
(1-r_f)^2 = \eta^2 \Rightarrow r_f = 1 \pm \eta \; .
\en
With $-1 \le \eta \le 0$, the integral is
\eq
(S_\ell, \> S_\ell) = i \frac{1}{\eta}
\int_{1 + \eta}^{1-\eta} dr_f \> r_f 
\left[ S_\ell \left( \partial_g \overline{S}_\ell \right) - \left(
\partial_g
S_\ell \right) \overline{S}_\ell \right]_\Sigma .
\en
where we have substituted as well
$\part_\eta f(g) = -( \eta/r_f) \part_g f(g)$ and $a^2(\eta)
= \eta^{-2}$.  In the region of integration, we also have
$\overline{S_\ell(g)} = S_\ell (-g)$.  Inside the integral, the
terms where the derivatives act
on $(g^2 -1)^{1/4}$ and its complex conjugate cancel out.
Defining $y = g/\sqrt{g^2 -1}$, 
\beqn
\nonumber S_\ell \left( \partial_\eta \overline{S}_\ell
 \right) - \left(\partial_\eta
S_\ell \right) \overline{S}_\ell  &=& |g^2 -1|^{1/2}\> \frac{d y}{dg}
\left[ P^{-\ell - 1/2}_{-3/2}(y)
\left( \partial_y 
P^{-\ell - 1/2}_{-3/2}(-y) \right) \right.  \\ 
& - & \left.
\left( \partial_y
P^{-\ell - 1/2}_{-3/2}(y) \right)
P^{-\ell - 1/2}_{-3/2}(-y) \right]
\eeqn
So we are left with the wronskian
of $P^{-\ell - 1/2}_{-3/2}(y)$ and  $P^{-\ell - 1/2}_{-3/2}(-y)$
times $dy(g)/dg$.
We can now use (\cite{gr}, equations 8.736.2, 8.334.3, 8.335.1,
and \cite{Absteg}, equation 8.1.8)
\beqn
\nonumber
 P^\mu_\nu(y) \frac{d}{dy} 
P^\mu_\nu(-y) - \frac{d}{dy} P^\mu_\nu(y) P^\mu_\nu(-y) &=&
-\frac{2}{\pi}\sin[(\mu + \nu)\pi] e^{-\mu \pi i}
[ P^\mu_\nu(y) \frac{d}{dy} 
Q^\mu_\nu(y) - \frac{d}{dy} P^\mu_\nu(y) Q^\mu_\nu(y)] \\
\nonumber &=& -\frac{2}{\pi}\sin[(\nu+\mu)\pi]
\frac {2^{2 \mu} \Gamma(\frac{\mu + \nu + 2}{2})
\Gamma(\frac{\mu + \nu + 1}{2})}
{(1 - y^2)
\Gamma(\frac{-\mu + \nu + 2}{2})
\Gamma(\frac{-\mu + \nu + 1}{2})} \\
\nonumber &=&
-\frac{1}{2^{2 \ell} \pi}\sin[-(\ell+2)\pi]
\frac {\Gamma(\frac{-\ell}{2})
\Gamma(\frac{-\ell -1}{2})}
{(1 - y^2)
\Gamma(\frac{\ell + 1 }{2})
\Gamma(\frac{\ell}{2})} \\
\nonumber &=&
-\frac{2}{(1-y^2)\pi}\sin[-(\ell+2)\pi]
\frac {\Gamma(-\ell - 1)}{\Gamma(\ell)}
\\
&=&
\frac{2}{(1-y^2)}\frac{1}{\Gamma(\ell) \Gamma(\ell +2)} \; .
\eeqn

Including 
\eq
\frac{d }{d g} y(g) = -\frac{1}{(g^2 -1)^{3/2}} \; ,
\en
using $|g^2 -1|^{1/2}/(g^2-1)^{1/2} = i$,
and substituting the above, the inner product becomes
\beqn
\nonumber   \frac{i}{\eta}
\int_{1 + \eta}^{1-\eta} dr_f \> r_f
\left[ S_\ell \left( \partial_g \overline{S}_\ell \right) - \left(
\partial_g
S_\ell \right) \overline{S}_\ell \right]_\Sigma 
\nonumber &=&
- \frac{2}{\Gamma(\ell) \Gamma(\ell +2)}
\frac{1}{\eta}\int_{1 + \eta}^{1-\eta} dr_f \> r_f \\
\nonumber &=& 
- \frac{2}{\Gamma(\ell) \Gamma(\ell +2)}\frac{1}{\eta}
\left[ \frac{(1-\eta)^2}{2} - \frac{(1+\eta)^2}{2} \right]
\\
&=&  \frac{4}{\Gamma(\ell) \Gamma(\ell +2)}
\eeqn
which is independent of $\eta$ as promised.  This also
shows that the supercurvature modes are properly normalized
for fixed time surfaces $-1 \le \eta \le 0$ in region $F$.
This suggests that at spatial infinity in the flat coordinates
the supercurvature modes have sufficiently fast falloff to
correspond to their inner product on a Cauchy surface.
(This is not true for fixed time surfaces
in region $R$ for example, as the supercurvature modes diverge
at spatial infinity.)

\subsection{Overlap along $r_c = 0$, for $\ell$ odd and $\ell = 2$}
In region $C$ the inner product
(equation (\ref{c1KG2})) is
\eq
(u,v) =
- i \cosh^2 r_c  \int_{-\pi/ 2}^{\pi / 2} 
d t_c \cos t_c \int d \Omega \left( {\part u \over \part r_c} 
\overline{v}
- u {\part \overline{v} \over \part r_c} \right)
\en
and will be taken on the Cauchy surface $r_c = 0$.  From their
definitions,
equations (\ref{c1xflat2}) and (\ref{c1xopen}),
the spatially-flat coordinates and the region $C$ coordinates
are related via
\beqn
\nonumber r_f &=&  \frac{\cos t_c \cosh r_c}{\sin t_c + \cos t_c \sinh 
r_c} \\
\eta &=& - \frac{1}{\sin t_c + \cos t_c \sinh r_c} \; ,
\label{flatc}
\eeqn
with $t_c >0$.
For $t_c <0$ we need the analytic continuation of the
basis function
\eq
\phi_{k \ell m} (x_f(r_c,t_c)) = \frac{i}
{\sqrt{\pi k}} \left( 1 + i k \eta (r_c,t_c)\right) e^{-ik \eta(r_c,t_c)} 
j_\ell (k r_f(r_c,t_c)) Y_{\ell m} (\Omega) .
\label{flatbasis3}
\en
As this basis function has no branch cuts as a function of
$r_f(r_c,t_c), \eta(r_c,t_c)$, the continuation to $t_c < 0$
is straightforward and the same for both positive and
negative frequency mode functions.

Using $\phi_{k \ell m}(x)$ to denote both the function in region $F$ and
its analytic continuation into region $t_c < 0$,
the inner product $\alpha_{k \ell m}$ between $\uel$ and $\phi$ is
\beqn
\nonumber  (\uel,\phi_{k\ell^\prime m^\prime}) &=& -i
\intp  d t_c \cos t_c  \int d \Omega\\
&\times&
\left[ u_{\ell m}(r_c,\Omega) \left(\part_{r_c} 
\overline{\phi}_{k \ell^\prime 
m^\prime}
(\eta,r_f,\Omega) \right)
- \left(\part_{r_c} u_{\ell m}(r_c,\Omega) \right) \overline{\phi}_{k
\ell^\prime 
m^\prime}
\right]|_{r_c = 0} \; .
\label{innerp}
\eeqn

For ease of calculation, we take
\beqn
\nonumber \phi_{k \ell m} (x) &=& \frac{i}{\sqrt{\pi k}} F_{k\ell}(x)
Y_{\ell m} (\Omega) \\
\uel (x_c) &=& \frac{1}{2}\sqrt{\Gamma(\ell)\Gamma 
(\ell + 2)} \> S_\ell(r_c)Y_{\ell m} (\Omega) 
\eeqn

The integral over $d \Omega$ is immediate, giving 
$\delta_{m m^\prime} \delta_{\ell \ell^\prime}$, and we suppress
these delta functions in the following. 
Pulling out normalization factors, the inner product becomes
\eq
\alpha_{k \ell m} =
- A(\ell,k)
\intp d t_c \cos t_c 
\left[ S_{\ell}(r_c) \left(\part_{r_c} 
\overline{F}_{k \ell}(\eta, r_f) \right)
- \left(\part_{r_c} S_{\ell}(r_c)  \right)\overline{F}_{k\ell}(\eta, r_f)
\right]|_{r_c = 0} \; ,
\en
where $A(\ell, k) \equiv  \frac{1}{2} \sqrt{\ell(\ell +1)}\Gamma(\ell) /
\sqrt{\pi k}$.

For $M=0$, $S_\ell(r_c)$ is independent of space $t_c$ and 
 can be pulled out of the integral  to give
\eq
(\uel,\phi_{k\ell m}) = 
- A(\ell,k)
\{ S_{\ell}(r_c) \part_{r_c}
- \part_{r_c} S_{\ell}(r_c)\}
\int_{-\pi / 2}^{\pi/2} 
d t_c \cos t_c  \overline{F}_{k\ell}(\eta, r_f)|_{r_c = 0} \; .
\en
Taking the limit $r_c \to 0$ (and remembering that $P^{-\ell -
1/2}_{-3/2}$
has imaginary argument) one has (\cite{Absteg}, equation 8.1.4)
\beqn
\nonumber S_{\ell}(0) &=&
\frac{P^{-\ell - 1/2}_{-3/2} (i \sinh r_c)}{\sqrt{i \cosh r_c}}\vert_{r_c
= 0} 
= e^{- i \pi  / 4} e^{ \pi i (\ell/2 + 1/4)} 
\frac{2^{-\ell - 1/2} \sqrt \pi }
{\Gamma({\ell+ 1 \over 2}) \Gamma({\ell + 3 \over 2})} \\
&\equiv &  e^{\pi i\ell/2 }  B(\ell)
\eeqn
and
\beqn
\nonumber \part_{r_c}S_\ell(r_c)|_{r_c = 0} 
&=& \part_{r_c}\left[ \frac{P^{- \ell - 1/2}_{-3/2} (i \sinh r_c) }
{ \sqrt{i \cosh r_c}} \right]_{r_c=0} \\
\nonumber &=& \frac{\ell P^{-\ell - 1/2}_{-1/2} (i \sinh r_c) }
{ (i \cosh r_c)^{3/2}}|_{r_c = 0}
 =\frac{e^{-3 i \pi / 4}
e^{\pi i(\ell/2 + 1/4)} 
\ell \>2^{-\ell - 1/2} \sqrt{\pi} }
{(\Gamma ({\ell +2 \over 2}))^2}\\
&\equiv&  e^{\pi i(\ell/2 - 1/2)} C(\ell) \; .
\eeqn
The inner product now becomes
\eq
\alpha_{k \ell m} = 
-i^{\ell}A(\ell,k)
\{B(\ell)  \part_{r_c}
+i C(\ell)\}
\int_{-\pi/2}^{\pi / 2} 
d t_c \cos t_c  \overline{F}_{k\ell}(\eta, r_f)|_{r_c = 0} \; ,
\en
where $A(\ell), B(\ell), C(\ell)$ are all real.

Again, both $\uel$ and $\phi_{k \ell m}$
are positive-frequency mode functions for the Bunch-Davies vacuum,
so that $\beta_{k \ell m}$ (defined in equation (\ref{alphabeta})) 
vanishes:
\beqn
\nonumber 0 &=& (\uel, \overline{\phi}_{k \ell^\prime m^\prime})\\
&=& 
-i^{\ell} A(\ell,k) (-1)^{m+1} \delta_{m^\prime, - m}
\{ B(\ell) \part_{r_c}
+ iC (\ell)\}
\intp
d t_c \cos t_c  F_{k \ell}(\eta, r_f)|_{r_c = 0} \; ,
\eeqn
The complex conjugate of this equation implies
\eq
B(\ell) \part_{r_c}\int_{-\pi / 2}^{\pi/2} 
d t_c \cos t_c  \overline{F}_{k \ell}(\eta, r_f)|_{r_c = 0}
=
i C(\ell)
\intp
d t_c \cos t_c  \overline{F}_{k \ell}(\eta, r_f)|_{r_c = 0} \; .
\en

Defining
\eq
\overline{\cal I}_\ell(k)  \equiv \int_0^{\pi / 2}
d t_c \cos t_c  \overline{F}_{k \ell}(\eta, r_f)|_{r_c = 0} \; 
\en
and using that $j_\ell(kr) = (-1)^\ell j_\ell(-kr)$ is real,
\beqn
\nonumber \overline{F}_{k \ell}(x_f (-t_c))\vert_{r_c = 0}
&=& \left(1 - i \frac{k}{\sin (-t_c)} \right) e^{i \frac{k}{\sin(-t_c)}}
j_\ell(k \cot(-t_c))
\\
&=&
\left(1 + i \frac{k}{\sin (t_c)} \right) e^{-i \frac{k}{\sin(t_c)}}
(-1)^\ell j_\ell(k \cot(t_c)) \\
\nonumber &=&(-1)^\ell F_{k \ell}(x_f (t_c))\vert_{r_c = 0} \; .
\eeqn
Thus we can write
\eq
\part_{r_c}\intp
d t_c \cos t_c  \overline{F}_{k \ell}(\eta, r_f)|_{r_c = 0}
= i  \frac{C(\ell) }{B(\ell)}
(\overline{\cal I}_\ell(k) +(-1)^\ell {\cal I}_\ell(k)) \; .
\en

Substituting in, the full inner product is thus
\beqn
\nonumber (\uel,\phi_{k \ell m}) &=& 
-2 i^{\ell +1} C(\ell) A(\ell,k) (\overline{\cal I}_\ell(k) +(-1)^\ell
{\cal I}_\ell(k))\\
\nonumber &=&  -i^{\ell +1} \frac{\ell \>2^{-\ell - 1/2} \sqrt{\pi} }
{(\Gamma ({\ell +2 \over 2}))^2} \sqrt{\ell(\ell +1)}
\frac{\Gamma(\ell)}{\sqrt{\pi k}} (\overline{\cal I}_\ell(k)+
(-1)^\ell {\cal I}_\ell(k))\\
\\
\nonumber &=&
- \frac{i^{\ell +1}}{\sqrt{k}} 
\frac{ 2^{-\ell - 1/2} \>\ell \> \sqrt{\ell(\ell +1)}\Gamma(\ell) }
{(\Gamma ({\ell +2 \over 2}))^2}
(\overline{\cal I}_\ell(k)+(-1)^\ell {\cal I}_\ell(k)) .
\label{pref}
\eeqn
%%%%%%
The calculation of the inner product thus requires the integral
\eq
{\cal I}_\ell(k) = \int_0^{\pi/2} 
d t_c \cos t_c (1 + i k \eta)e^{-i k \eta}j_\ell(k r_f)  |_{r_c = 0} 
\; .
\en
Changing coordinates and expressing
$\eta, t_c$ in terms of $r_f$ (there is only one free coordinate
as $r_c$ has been fixed to zero), using
\eq
\eta = - \sqrt{r_f^2 +1} \; , \; \; \tan t_c = \frac{1}{r_f} ,
\en
gives
\eq
{\cal I}_\ell(k) =
 \int_0^{\infty} d r_f \frac{r_f }{ (r_f^2 +1)^{3 / 2}}
\left(1 - i k \sqrt{r_f^2 + 1} \right)e^{i k \sqrt{r_f^2 +1}} j_\ell(k
r_f) \; .
\en
To proceed, note
\eq
\frac{r_f }{ (r_f^2 +1)^{3 / 2}}
\left(1 - i k \sqrt{r_f^2 + 1} \right)e^{i k \sqrt{r_f^2 +1}}
= -\part_{r_f} (\frac{e^{ik \sqrt{r_f^2 +1}}}{\sqrt{r_f^2 +1}} ) \; .
\en
For $\ell$ odd, we need the imaginary part of ${\cal I}_\ell(k)$ and
so can use (\cite{Absteg}, equation 10.1.45)
\eq
\frac{i \sin{k \sqrt{r_f^2 +1}}}{k\sqrt{r_f^2 +1}}
= i\sum_{n=0}^\infty (2 n +1) j_n (k r_f) j_n (k) P_n(0) 
\en
Consequently, for $\ell$ odd the integral of interest is
\eq
{\rm Im} \; {\cal I}_{\ell, odd}(k) =
-ik \sum_{n = 0}^\infty (2n +1) j_n(k) P_n(0)
\int_0^\infty dr_f \part_{r_f}(j_n(k r_f)) j_\ell (kr_f)
\label{iell}
\en
Using Mathematica,
\beqn
\nonumber \int_0^\infty dr \part_r j_n (kr_f) j_\ell(kr_f)
&=& \frac{\ell(\ell + 1) - n(n+1)}{(\ell -n - 1)(\ell - n +1)
(\ell + n)(2 + \ell + n)} \cos [\pi \frac{\ell - n}{2}] \\
(\ell \; {\rm odd}) \; 
&=&\frac{\ell(\ell + 1) - n(n+1)}{(\ell -n - 1)(\ell - n +1)
(\ell + n)(2 + \ell + n)} \sin \frac{\pi \ell}{2} \sin \frac{\pi n}{2}
\eeqn
where it appears that 
$n$ odd is required as well.
However, looking at the sum (equation \ref{iell}), we see this term is
multiplied by $P_n(0)$ which vanishes for $n$ odd.  
So the only possibility for a nonzero 
term is if the denominator in this expression vanishes, that is if
$\ell = n \pm 1$.  Substituting in these values and the
definition of the Legendre polynomials we get
\beqn
\nonumber {\rm Im} \; {\cal I}_{\ell , odd}(k) 
&=&- ik \frac{- \ell}{2 \ell + 1} \frac{\pi}{2}
P_{\ell - 1}(0) (j_{\ell-1}(k) + j_{\ell +1}(k)) \\
&=& i \ell \frac{\pi}{2}
P_{\ell - 1}(0) j_\ell(k)
\eeqn

Including the prefactors (equation (\ref{pref})) and
after some algebra with $\Gamma$ functions one gets that
the overlap for $\ell$ odd is
\eq
\alpha_{k \ell m} = -i \sqrt{2 \ell(\ell +1)} k^{-1/2}j_\ell(k) \; \; ,
\ell \; {\rm odd}
\en
agreeing with equation (\ref{result}).

For $\ell$ even the calculation is more involved because the
identity required in this case is (\cite{Absteg}, equation 10.1.46)
\eq
\frac{\cos {k \sqrt{r_f^2 +1}}}{k\sqrt{r_f^2 +1}} 
=
\begin{array}{ll} 
-\sum_{n=0}^\infty (2 n +1) y_n (k) j_n (k r_f) P_n(0)  
\; ,& \; r_f < 1 \\
 \; & \\
-\sum_{n=0}^\infty (2 n +1) y_n (k r_f) j_n (k ) P_n(0)  
\; , &\; r_f>1 \; .
\end{array}
\label{expsum}
\en
so that the range of integration in ${\cal I}_\ell(k)$
is split from $0 \le r_f \le 1$ and $1 \le r_f \le \infty$.
As the resummation over $\alpha_{k \ell m} \phi_{k \ell m}$ in
equation (\ref{alphacheck}) works
equally well for $\ell$ odd and even, it does not seem enlightening
to pursue the calculation for even $\ell$ in all generality here.

We can check a specific case, $\ell = 2$, by integrating by
parts and using the definition of $j_2 (k r_f)$,
\eq
j_2(k{r_f}) = (k{r_f})^2 \left(-{1 \over 
k^2 {r_f}} \part_{r_f} \right)^2 {\sin k{r_f} 
\over k {r_f}}
\; .
\label{besj}
\en
to get
\beqn
\nonumber {\cal I}_2(k) 
&=& \int_0^\infty dr_f \frac{r_f  \left(1- i k \sqrt{r_f^2 
+1} \right)}{ (r_f^2 +1)^{3/2}} e^{i k \sqrt{r_f^2 +1}}
 (kr_f)^2 \left(- \frac{1 }{ k^2 r_f} \part_{r_f} \right)^2 \frac{\sin
kr_f }{ k r_f} \\
\nonumber &=& -\lim_{r_f \to 0}
k^{-3} \frac{r_f  \left(1- i k \sqrt{r_f^2 +1} \right)}
{(r_f^2 +1)^{3/2}} e^{i k \sqrt{r_f^2 +1}} \part_{r_f}
\frac{\sin kr_f }{r_f} \\
\nonumber &-& k^{-3} 
\frac{1 }{ r_f^2}\part_{r_f} \left[ \frac{r_f^2 \left(1- i k \sqrt{r_f^2
+1} 
\right)}{(r_f^2 +1)^{3/2}} e^{i k \sqrt{r_f^2 +1}} \right]
{\sin kr_f}|^\infty_0 \\
&+& k^{-3} \int_0^\infty dr_f \frac{\sin k r_f }{ r_f} \part_{r_f} \left(
\frac{1 }{ r_f}\part_{r_f} \left[ \frac{r_f^2 \left(1- i k \sqrt{r_f^2
+1} \right)}
{(r_f^2 +1)^{3/2}} e^{i k \sqrt{r_f^2 +1}} \right]\right) \; .
\eeqn 
The only nonzero boundary term is the second one, at $r_f = 0$, which
can be read off, as it is only nonzero when the derivatives
act on $r_f$ rather than on $\sqrt{r_f^2 +1}$.
Substituting $r_f \to \mp (x - x^{-1})/2 $ 
and using Mathematica
gives
\eq
{\cal I}_2(k)= {6i(e^{ik} - 1) \over k^3} +{6 e^{ik} \over k^2} - 
i{(1 + 2 e^{ik})\over k} \; .
\en
Combining the integral with the prefactors (eqn. (\ref{pref})),
and taking the real part gives
\eq
\alpha_{k, \ell =  2, m} = 
-i 2\sqrt{ \frac{3}{k}} \frac{1}{k^3} 
\left[ 3 \sin k - 3 k \cos k -  k^2 \sin k \right] = -2i \sqrt{3} \>
k^{-1/2} \> j_2 (k),
\en
again in agreement with equation (\ref{result}).

\subsection{Overlap along $\eta = 0$, for $\ell = 1$}
Here we discuss the integral (equation (\ref{I_ell}))
\eq
I_\ell (k)_{\vert \Sigma} = \int_0^\infty dr_f \left[ r_f^2 \>
j_\ell (k r_f) S_\ell (r_r) \right]_\Sigma 
\en
evaluated on the surface in region $F$ corresponding to $\eta = 0$.
For $\eta = 0$, $g = (1 + r_f^2)/(2 r_f)$ and so 
\eq
I_\ell(k)|_{\Sigma} =
\int_0^\infty d{r_f} {r_f}^2 j_\ell( k {r_f}) (g^2 - 1)
\part_g Q_\ell(g)
\en
becomes for $\ell = 1$
\eq
I_1(k)|_{\eta = 0} =
-\sqrt{\frac{2}{\pi}}\frac{1}{\Gamma(3)}
\int_0^\infty  d{r_f} \> {r_f}^2 \left[\frac{\sin k {r_f}}{(k{r_f})^2} -
\frac{\cos k {r_f}}{k{r_f}}
\right] \left( \frac{(1 - {r_f}^2)^2}{8{r_f}^2} \ln
\left(\frac{(1+{r_f})^2}{(1-{r_f})^2} \right)
- \frac{1 + {r_f}^2}{2 {r_f}} \right)
\en
This can be rewritten as
\eq
I_1(k)|_{\eta = 0} =
-\sqrt{\frac{1}{2\pi}}
\int_0^\infty d{r_f}  \left[\frac{\sin k {r_f}}{(k{r_f})^2} - 
\frac{\cos k {r_f}}{k{r_f}}
\right] \left(\frac{(1 - {r_f}^2)^2}{4} 
\ln \left| \frac{1+{r_f}}{1-{r_f}} \right|
- {r_f}\frac{1 + {r_f}^2}{2 } \right)
\en
and integrated using Mathematica by considering
\beqn
\int_0^1 d{r_f}   \left[\frac{\sin k {r_f}}{(k{r_f})^2} - \frac{\cos k
{r_f}}{k{r_f}}
\right] \left(\frac{(1 - {r_f}^2)^2}{4} \ln \left(
\frac{1+{r_f}}{1-{r_f}} \right) 
- {r_f}\frac{1 + {r_f}^2}{2 } \right)
& & \\
+
\int_1^\infty d{r_f}  \left[\frac{\sin k {r_f}}{(k{r_f})^2} -
 \frac{\cos k {r_f}}{k{r_f}}
\right] \left( \frac{(1 - {r_f}^2)^2}{4} \ln \left( \frac{1+{r_f}}{-1
+{r_f}} \right)
- {r_f}\frac{1 + {r_f}^2}{2 } \right)
& & \\
\eeqn
to get
\beqn
\nonumber I_1(k ) &=& -\sqrt{\frac{1}{2\pi}}
\{ \lim_{{r_f} \to \infty}{4 \over 3 k^2} \sin k {r_f}  + 
\frac{2}{k^5} \left[ -(\ci(\infty) - \ci(-\infty))(\cos(k) + k \sin k)
\right. 
\\
\nonumber
&+& \left. (\Si(\infty) - \Si(-\infty))(-k \cos k + \sin k) \right]
+ \frac{2 i \pi}{k^5}(\cos k + k \sin k) \} \\
&=& 
-\sqrt{\frac{1}{ 2\pi}} \left(
\lim_{{r_f} \to \infty}{4 \over 3 k^2} \sin k {r_f} +
-\frac{2 \pi}{k^5} (\sin k - k \cos k) \right) .
\eeqn
This gives
\eq
\alpha_{k, \ell = 1, m} =
i \frac{k^3}{\sqrt{\pi k}} \left(
\lim_{{r_f} \to \infty}{4 \over 3 k^2} \sin k {r_f} +
-\frac{2 \pi}{k^5} (\sin k - k \cos k) \right) =
-2 i k^{-1/2} \> j_1 (k) ,
\en
where the oscillating first term has been dropped as it
has the wrong asymptotics as $k \to \infty$.
(If $\lim_{r_f \to \infty} \sin k r_f$ was finite rather than the
giving zero in this limit, the overlap 
would diverge as $k \to \infty$.)
With this, $\alpha_{k\ell m}$ again agrees with the
results found above.

\section*{Acknowledgments}
D.K. thanks A. Guth, and
J.D.C. thanks A. Anderson, S. Axelrod, L. Ford, D.E. Freed, A. Kent,
and A. Vilenkin for conversations and is especially grateful to
M. White.  J.D.C. also thanks the Aspen Center for Physics,
A. Loeb of the Harvard-Smithsonian Center for Astrophyics, the
Tufts Institute of Cosmology and
the Insitute d'Astrophysique de Paris for hospitality in the course
of this work.
The work of D. K. has been supported in part by NSF PHY-92-18167.
J. D. C. is supported by an NSF Career Advancement Award, NSF
PHY-9722787, and, at the commencement of this work, 
by an ONR grant as a Mary Ingraham Bunting
Institute Science Scholar at Harvard University.


\begin{thebibliography}{9999}
\bibitem{birdavies}  N. D. Birrell and P. C. W. Davies, {\it Quantum
Fields in Curved Space} (Cambridge University Press, New York, 1982).
\bibitem{sfulling}  S. A. Fulling, {\it Aspects of Quantum Field
Theory in 
Curved Spacetime} (Cambridge University Press, New York, 1989).
\bibitem{KolbTurner}  A. H. Guth, Phys. Rev. D {\bf 23}, 347 (1981).
See also E. W. Kolb and M. S. Turner, {\it The Early 
Universe} (Addison-Wesley, New York, 1990); A. D. Linde, {\it Particle 
Physics and Inflationary Cosmology} (Harwood, Chur, 1990).
\bibitem{bgt}  M. Bucher, A. Goldhaber, and
N. Turok, Phys. Rev. D {\bf 52}, 3314 (1995);
Nucl. Phys. Proc. Suppl. {\bf 43} (1995) 173 .
\bibitem{openi} K. Yamamoto, M. Sasaki, and T. 
Tanaka, Astrophys. J. {\bf 455}, 412 (1993); A. D. Linde, Phys. Lett.
{\bf B351}, 99 
(1995); A. D. Linde and A. Mezhlumian, Phys. Rev. D {\bf 52}, 6789
(1995).
\bibitem{colegott}  S. Coleman and F. de Luccia, Phys. Rev. D {\bf
21},
3305 (1980); J. R. Gott, Nature {\bf 295}, 304 (1982); J. R. Gott and T.
S. Statler, Phys. Lett. {\bf B136}, 157 (1984).
\bibitem{lythwos} D. H. Lyth and A. Woszczyna, Phys. Rev. D {\bf 52},
3338 (1995); J. Garcia-Bellido, A. R. Liddle, D. H. Lyth, and D. Wands,
Phys. Rev. D {\bf 52}, 6750 (1995).
\bibitem{sasaki95}  M. Sasaki, T. Tanaka, and K. Yamamoto, Phys. Rev.
D {\bf 51}, 2979 (1995).
\bibitem{Mosch} U. Moschella and R. Schaeffer, \lq\lq Quantum 
Fluctuations in an Open Universe," preprint gr-qc/9707007.
\bibitem{desinfra}  A. D. Linde, Phys. Lett. {\bf B116}, 335 (1982);
A. A. Starobinsky, Phys. Lett. {\bf B117}, 175 (1982); A. Vilenkin and L.
H. Ford, Phys. Rev. D {\bf 26}, 1231 (1982); B. Allen, Phys. Rev. D {\bf
32}, 3136 (1985); L. H. Ford and A. Vilenkin, Phys. Rev. D {\bf 33}, 2833
(1986); B. Allen and A. Folacci, Phys. Rev. D {\bf 35}, 3771 (1987); D.
Polarski, Phys. Rev. D {\bf 43}, 1892 (1991).
\bibitem{kg93} K. Kirsten and J. Garriga, Phys. Rev. D {\bf 48}, 567
(1993).
\bibitem{sccmb} A. D. Linde and A. Mezhlumian in reference 5 above,
J. Garcia-Bellido, Phys. Rev. D {\bf 54}, 2473 (1996)
\bibitem{garriga} J. Garriga, Phys. Rev. D {\bf 54}, 4764 (1996).
\bibitem{sasaki96} K. Yamamoto, M. Sasaki, and T. Tanaka, Phys. Rev. 
{\bf 54}, 5031 (1996); M. Sasaki, T. Tanaka, 
Phys.Rev. D {\bf 54} 4705 (1996).
\bibitem{gar-97} J. Garcia-Bellido, J. Garriga, X. Montes,
``Quasi-Open Inflation,'' hep-ph/9711214.
\bibitem{dkpre} D. I. Kaiser, in preparation.
\bibitem{newreh} L. Kofman, A. Linde, and A. A. Starobinsky, Phys. Rev.
Lett. {\bf 73}, 3195 (1994); Y. Shtanov, J. Traschen, and R.
Brandenberger, Phys. Rev. D {\bf 51}, 5438 (1995); D. Boyanovsky {\it et
al.}, Phys. Rev. D {\bf 51}, 4419 (1995), {\bf 52}, 6805 (1995), and {\bf
54}, 7570 (1996); M. Yoshimura, Prog. Theo. Phys. {\bf 94}, 8873 (1995);
D. I. Kaiser, Phys. Rev. D {\bf 53}, 1776 (1996).
\bibitem{jdcdk}  J. D. Cohn and D. I. Kaiser, in preparation.
\bibitem{rhb} V. Mukhanov, H. Feldman, and R. Brandenberger, Phys. 
Rep. {\bf 215}, 203 (1992); V. Mukhanov, L. Abramo, and R. Brandenberger, 
Phys. Rev. Lett. {\bf 78}, 1624 (1997) and Phys. Rev. D {\bf 56}, 3248
(1997); See also M. Sasaki, T. Tanaka, ``Super-Horizon Scale Dynamics
of Multi-Scalar Inflation,'' gr-qc/9801017.
\bibitem{ballen}  B. Allen, Phys. Rev. D {\bf 51}, 3136 (1985).
\bibitem{hawkellis} S. W. Hawking and G. F. R. Ellis, {\it
Large-Scale 
Structure of Spacetime} (Cambridge University Press, New York, 1973).
\bibitem{bunchdav} T. S. Bunch and P. C. W. Davies, Proc. Roy. Soc. A 
{\bf 360}, 117 (1978); T. S. Bunch and P. C. W. Davies, J. Phys. A {\bf
11}, 1315 
(1978).  See also R. H. Brandenberger, Nucl. Phys. {\bf B245}, 328
(1984); A. H. 
Guth and S.-Y. Pi, Phys. Rev. D {\bf 32}, 1899 (1985).
\bibitem{buchtur95} M. Bucher and N. Turok, Phys. Rev. D {\bf 52},
5538 (1995).
\bibitem{dewitt}  B. S. DeWitt, in {\it Relativity, Groups, and
Topology II, Les Houches 1983}, ed. B. S. DeWitt and R. Stora
(North-Holland, New York, 1984).
\bibitem{Absteg} M. Abramowitz and I. Stegun, {\it Handbook of 
Mathematical Functions} (Dover, New York, 1965).
\bibitem{wallb} See for example T. Hamazaki, M. Sasaki,
T. Tanaka, K. Yamamoto, Phys. Rev. D {\bf 53}, 2045 (1996);
J. Garcia-Bellido, Phys. Rev. D {\bf 54}, 2473 (1996);
J. Garriga, Phys. Rev. D {\bf 54} 4764 (1996);
J. Garriga, X. Montes, M. Sasaki, T. Tanaka,
``Canonical Quantization of Cosmological Perturbations
in the One Bubble Open Universe,'' astro-ph/9706229.
\bibitem{rajaraman} R. Rajaraman, {\it Solitons and Instantons}
(North-Holland, New York, 1982).
\bibitem{gr} I. S. Gradshteyn, I. M. Ryzhik,
{\it Table of Integrals, Series and Products}, Fourth Edition (Academic
Press, Inc., San Diego, 1980).

\end{thebibliography}
\end{document}